\documentclass[a4paper,twoside]{article}
\usepackage[authoryear]{natbib}
\bibpunct{(}{)}{;}{a}{}{,}
\usepackage{graphicx}
\usepackage{psfig}
\date{} %Please leave the date blank
\newcommand{\kms}{\mbox{km\,s$^{-1}$}}
\newcommand{\affil}[1]{$^{\rm #1}$}
\newcommand{\hi}{H\,\textsc{i}}
\newcommand{\HI}{H\,{\sc i}}
\newcommand{\HII}{H\,{\sc ii}}
\newcommand{\hii}{H\,\textsc{ii}}
\newcommand{\tel}{ASKAP}
\def\lesssim{\mathrel{\hbox{\rlap{\hbox{\lower4pt\hbox{$\sim$}}}\hbox{$<$}}}}
\def\gtrsim{\mathrel{\hbox{\rlap{\hbox{\lower4pt\hbox{$\sim$}}}\hbox{$>$}}}}
\title{\large\bf\flushleft Science With The Australian Square Kilometre Array Pathfinder}
\author{\parbox{\textwidth}{\flushleft
\vspace{-0.5cm}
{\it S. Johnston\affil{A,X},
M. Bailes\affil{B}, N. Bartel\affil{C}, C. Baugh\affil{D},
M. Bietenholz\affil{C,W}, C. Blake\affil{B}, R. Braun\affil{A}, J. Brown\affil{E},
S. Chatterjee\affil{F},
J. Darling\affil{G}, A. Deller\affil{B}, R. Dodson\affil{H},
P. G. Edwards\affil{A},  R. Ekers\affil{A}, S. Ellingsen\affil{I},
I. Feain\affil{A},
B. M. Gaensler\affil{F},
M. Haverkorn\affil{J}, G. Hobbs\affil{A}, A. Hopkins\affil{F},
C. Jackson\affil{A}, C. James\affil{K}, G. Joncas\affil{L},
V. Kaspi\affil{M}, V. Kilborn\affil{B}, B. Koribalski\affil{A}, R. Kothes\affil{E},
T. L. Landecker\affil{N}, E. Lenc\affil{B}, J. Lovell\affil{I}, 
J.-P. Macquart\affil{O}, R. Manchester\affil{A}, D. Matthews\affil{P},
N. M. McClure-Griffiths\affil{A},
R. Norris\affil{A},
U.-L. Pen\affil{Q}, C. Phillips\affil{A}, C. Power\affil{B}, R. Protheroe\affil{K},
E. Sadler\affil{F}, B. Schmidt\affil{R}, I. Stairs\affil{S},
L. Staveley-Smith\affil{T}, J. Stil\affil{E},
R. Taylor\affil{E}, S. Tingay\affil{U}, A. Tzioumis\affil{A},
M. Walker\affil{V}, J. Wall\affil{S}, M. Wolleben\affil{N}
}\\
\vspace{0.4cm}
{\tiny \affil{A}\,Australia Telescope National Facility, CSIRO, PO Box 76, Epping, NSW 1710, Australia}\\
{\tiny \affil{B}\,Centre for Astrophysics and Supercomputing, Swinburne University of Technology, PO Box 218, Hawthorn, Vic 3122, Australia}\\
{\tiny \affil{C}\,Dept of Physics and Astronomy, York University, Toronto, ON M3J 1P3, Canada}\\
{\tiny \affil{D}\,Institute for Computational Cosmology, University of Durham, Durham, DH1 3LE, UK}\\
{\tiny \affil{E}\,Dept of Physics and Astronomy, University of Calgary, Calgary, AB T2N 1N4, Canada}\\
{\tiny \affil{F}\,School of Physics, The University of Sydney, NSW 2006, Australia}\\
{\tiny \affil{G}\,Center for Astrophysics and Space Astronomy, University of Colorado, 389 UCB, Boulder, CO 80309-0389, USA}\\
{\tiny \affil{H}\,Observatorio Astronomico Nacional, Alcara de Henares, Spain}\\
{\tiny \affil{I}\,School of Mathematics and Physics, University of Tasmania, Private Bag 21, Hobart, Tas 7001, Australia}\\
{\tiny \affil{J}\,NRAO Jansky Fellow: Astronomy Dept, University of California-Berkeley, Berkeley, CA 94720, USA}\\
{\tiny \affil{K}\,School of Chemistry \& Physics, University of Adelaide, SA 5006, Australia}\\
{\tiny \affil{L}\,Dept de Physique et Observatoire du Mont Megantic, Universite Laval, Quebec, QC G1K 7P4, Canada}\\
{\tiny \affil{M}\,Dept of Physics, McGill Unviersity, Montreal, QC H3A 2T8, Canada}\\
{\tiny \affil{N}\,Dominion Radio Astrophysical Observatory, Herzberg Institute of Astrophysics, NRC, Penticton, BC, Canada}\\
{\tiny \affil{O}\,NRAO Jansky Fellow: Astronomy Dept, California Institute of Technology, Pasadena, CA 91125, USA}\\
{\tiny \affil{P}\,Dept of Physics, La Trobe University, Vic 3086, Australia}\\
{\tiny \affil{Q}\,Canadian Insititute for Theoretical Astrophysics, University of Toronto, Toronto, ON M5S 3H8, Canada}\\
{\tiny \affil{R}\,Mount Stromlo and Siding Spring Observatory, Private Bag, Weston Creek, Canberra, ACT 2601, Australia}\\
{\tiny \affil{S}\,Dept of Physics and Astronomy, University of British Columbia,6224 Agricultural Road, Vancouver, BC V6T 1Z1, Canada}\\
{\tiny \affil{T}\,School of Physics, University of Western Australia, Crawley, WA 6009, Australia}\\
{\tiny \affil{U}\,Dept of Imaging and Applied Physics, Curtin University of Technology, Bentley, WA, Australia}\\
{\tiny \affil{V}\,Manly Astrophysics Workshop Pty Ltd, Manly, NSW 2095, Australia}\\
{\tiny \affil{W}\,Hartebeesthoek Radio Observatory, PO Box 443, Krugersdorp 1740, South Africa
}\\
{\tiny \affil{X}\,Email: Simon.Johnston@csiro.au}}}
\begin{document}
\twocolumn[
\begin{changemargin}{.8cm}{.5cm}
\begin{minipage}{.9\textwidth}
\vspace{-1cm}
\maketitle
%
%
%%%%%%%%%%%%%     ABSTRACT    %%%%%%%%%%%%%
%sof no more than 200 words here.
\small{\bf Abstract:}
The future of cm and m-wave astronomy lies with the
Square Kilometre Array (SKA),
a telescope under development by a consortium of 17 countries that will be 50
times more sensitive than any existing radio facility.
Most of the key science for the SKA will be addressed through large-area
imaging of the Universe at frequencies from a few hundred MHz to a few GHz.
The Australian SKA Pathfinder (\tel) is a technology demonstrator aimed
in the mid-frequency range, and achieves instantaneous wide-area imaging
through the development and deployment of phased-array feed systems on
parabolic reflectors.
The large field-of-view makes \tel\ an unprecedented synoptic
telescope that will make substantial advances in SKA key science.
\tel\ will be located at the Murchison Radio 
Observatory in inland Western Australia, one of
the most radio-quiet locations on the Earth and one of
two sites selected by the international community as a potential
location for the SKA.
In this paper, we outline an ambitious science program for \tel,
examining key science such as understanding the evolution, formation
and population of galaxies including our own, understanding the
magnetic Universe, revealing the transient radio sky and searching
for gravitational waves.

%%%%%%%%%%%%%     KEYWORDS    %%%%%%%%%%%%%
\medskip{\bf Keywords:} telescopes
% Please write all keywords in lower case. PASA uses the
% standard list of subject headings adopted by The Astrophysical Journal
% and available from http://www.journals.uchicago.edu/ApJ/keywords_text.html.
% Keywords are separated by em-dashes, i.e. ---

%%%%%%%%DO NOT EDIT%%%%%%%%%%%%
\medskip
\medskip
\end{minipage}
\end{changemargin}
]
\small
%%%%%%%%EDIT FROM HERE%%%%%%%%%%%%
\section{Introduction}
\begin{table*}
\begin{center}
\caption{System parameters for \tel}
\begin{tabular}{lccc}
\hline & \vspace{-3mm} \\
\multicolumn{1}{c}{Parameter} & Symbol & Strawman & Expansion \\
\hline & \vspace{-3mm} \\
Number of Dishes & $N$ & 30 & 45 \\
Dish Diameter (m) & & 12 & 12\\
Dish area (m$^2$) & $A$ & 113 & 113\\
Total collecting area (m$^2$) & & 3393 & 5089\\
Aperture Efficiency & $\epsilon_a$ & 0.8 & 0.8 \\
System Temperature (K) & $T$ & 50 & 35 \\
Number of beams & & 30 & 30\\
Field-of-view (deg$^2$) & $F$ & 30 & 30 \\
Frequency range (MHz) & & 700 $-$ 1800 & 700 $-$ 1800\\
Instantaneous Bandwidth (MHz) & $B$ & 300 & 300\\
Maximum number of channels & & 16000 & 16000 \\
Maximum baseline (m) & & 2000 & 400, 8000\\
\hline & \vspace{-3mm} \\
\end{tabular}
\label{intro:params}
\end{center}
\end{table*}
The Australian SKA Pathfinder (\tel) is a next generation radio
telescope on the strategic
pathway towards the staged development of the Square Kilometre Array
(SKA). The \tel\ project is international
in scope and includes partners in Australia, Canada, the Netherlands and South Africa.
This paper, which concentrates on the science made possible with \tel\
was written jointly between Australian and Canadian
research scientists following a collaborative agreement signed between
the CSIRO, Australia and the National Research Council of Canada.
\\

\noindent \tel\ has three main goals:
\begin{itemize}
\item to carry out world-class, ground breaking observations directly
relevant to the SKA Key Science Projects;
\item to demonstrate and prototype technologies for the mid-frequency
SKA, including field-of-view enhancement by focal-plane phased-arrays
on new-technology 12-m class parabolic reflectors;
\item to establish a site for radio astronomy in Western Australia where
observations can be carried out free from the harmful effects of 
radio interference.
\end{itemize}

Following international science meetings held in April
2005 and March 2007, seven main science themes have been identified 
for \tel. These are
extragalactic \HI\ science, continuum science, polarization science,
Galactic and Magellanic science, VLBI science, pulsar science and
the radio transient sky. In this paper, we first give a general
science overview, then outline the
system parameters for \tel, before allocating a section to a summary
of each science theme. A larger, and more complete, version of the 
science case for \tel\ will be published elsewhere.

\subsection{\tel\ and SKA science}
The SKA will impact a wide range of science from fundamental
physics to cosmology and astrobiology. The SKA Science Case ``Science
with the Square Kilometre Array" (Carilli \& Rawlings 2004) identifies
compelling questions that will be addressed as key science by the SKA :
\begin{itemize}
\item understanding the cradle of life by imaging the environments of the
formation of earth-like planets, the precursors to biological molecules,
and carrying out an ultra-sensitive search for  evidence of
extra-terrestrial intelligence,
\item carrying out fundamental tests of the theory of gravity by
using radio waves to measure the strong space-time  warp of pulsars and
black holes and timing of arrays of pulsars over large
areas of the sky to detect long-wavelength gravitational
waves propagating through the Galaxy,
\item tracing the origin and evolution of cosmic magnetism by
measuring the properties of polarized radio waves from galaxies over
cosmic history
\item charting the cosmic evolution of galaxies and large-scale structure,
the cosmological properties of the universe and dark energy, the
imaging of atomic hydrogen emission from galaxies and the
cosmic web from the present to time of the first galaxies, and
\item probing the dark ages and the epoch of reionization of the
Universe when the first compact sources of energy emerged.
\end{itemize}

The technological innovation of \tel\ and the unique radio quiet location in
Western Australia will enable a powerful synoptic survey instrument that
will  make substantial advances in SKA technologies and on three of 
the SKA key science projects:
the origin and evolution of cosmic magnetism, the evolution of
galaxies and large scale structure, and strong field tests of gravity.
The headline science goals for \tel\ are:
\begin{itemize}
\item The detection of a million galaxies in atomic hydrogen emission
across 80\% of the sky out to a redshift of 0.2 to understand galaxy
formation and gas evolution in the nearby Universe.
\item The detection of synchrotron radiation from 60 million galaxies to
determine the evolution, formation and population of galaxies across
cosmic time and enabling key cosmological tests.
\item The detection of polarized radiation from over 500,000 galaxies,
allowing a grid of rotation measures at $10'$  to explore
the evolution of magnetic fields in galaxies over cosmic time.
\item The understanding of the evolution of the interstellar medium of
our own Galaxy and the processes that drive its chemical and physical
evolution.
\item The characterization of the radio transient sky through
detection and monitoring of transient sources such as gamma ray
bursts, radio supernovae and intra-day variables.
\item The discovery and timing of up to 1000 new radio pulsars to find
exotic objects and to pursue the direct detection of gravitational waves.
\item The high-resolution imaging of intense, energetic phenomena
through improvements in the Australian and global Very Long Baseline networks.
\end{itemize}

\subsection{System Parameters}
Table~\ref{intro:params} gives the \tel\ system parameters.
The first column gives the parameter with the second column listing
the symbol used in the equations in this section.
The strawman (or base model) parameters for \tel\ are given in 
the third column of Table~\ref{intro:params} and these strawman 
assumptions have been used throughout this paper.
Likely upgrade or expansion paths include the addition of further
dishes and/or the cooling of the focal plane array elements to
provide a lower system temperature. Parameters for this expansion
path are listed in the column 4 of Table~\ref{intro:params}.

\tel\ is designed to be a fast survey telescope. A key metric in
this sense is the survey speed expressed in the number of square 
degrees per hour that the sky can be surveyed to a given sensitivity.
Survey speeds and sensitivity for an interferometer
like \tel\ have been derived elsewhere (e.g. Johnston \& Gray 2006)
and the full derivation will not be shown here.
To summarise, the time, $t$, required to reach a given sensitivity
limit for point sources, $\sigma_s$, is
\begin{equation}
t = \left( \frac{2\,\,\, k\,\,\, T}{A\,\,\, N\,\,\, \epsilon_a \,\,\, \epsilon_c
} \right)^2 \frac{1}{\sigma_s^2\,\,\, B\,\,\, n_p}
\end{equation}
where $B$ is the bandwidth (Hz), $n_p$ the number of polarizations,
$A$ is the collecting area of a single element (m$^2$), $N$ is the number of
elements and $\epsilon_a$ and $\epsilon_c$
represent dish and correlator efficiencies. The system temperature is
$T$ with $k$ being the Boltzmann constant.
The number of square degrees per second that
can be surveyed to this sensitivity limit is
\begin{equation}
SS_s = F\,\,\,B\,\,\,n_p \left( \frac{A\,\,\, N\,\,\, \epsilon_a \,\,\, \epsilon_c \,\,\, \sigma_s}{2\,\,\,k\,\,\, T} \right)^2
\label{eq:ss}
\end{equation}
where $F$ is the field of view in square degrees.
The surface brightness temperature survey speed is given by
\begin{equation}
SS_t = F \,\,\,B\,\,\,n_p  \left( \frac{\epsilon_c \,\,\, \sigma_t}{T} \right)^2 \,\,\, f^2  \,\,\, \epsilon_s^{-2}
\label{eq:sb}
\end{equation}
where now $\sigma_t$ denotes the sensitivity limit in K and $f$ relates to
the filling factor of the array via
\begin{equation}
f = \frac{A \,\,\, \epsilon_a \,\,\, N \,\,\, \Omega\,\,\, \epsilon_s}{\lambda^2}
\label{eq:ff}
\end{equation}
Here, $\epsilon_s$ is a `synthesised aperture efficiency' which is related 
to the weighting of the visibilities and is always $\leq 1$.

There is interplay between these parameters when trying to maximise
the survey speed for a given expenditure. For the majority of the science 
that will be considered here, the value of $SS_s$ and $SS_t$ are 
critical parameters, although the instantaneous sensitivity is also important,
especially for pulsar science.
Although these equations are useful, they are not the entire story.
For example, the effects of good (u,v) coverage on the image quality
and dynamic range do not appear in the equations. Furthermore,
one should also not neglect the total bandwidth available for a 
spectral line survey.
If the total bandwidth (or velocity coverage) is insufficient to 
cover the required bandwidth (velocity range)
of a given survey, the survey speed suffers as a result of having to
repeat the same sky with a different frequency setting.

In Table~\ref{intro:speed} we list values of the sensitivity and
survey speed for different `typical' surveys for both the strawman
and the expansion parameters. The first entry gives a continuum survey
where the entire 300~MHz of bandwidth is exploited and a desired
1-$\sigma$ sensitivity of 100~$\mu$Jy is required. The second entry gives a
spectral line survey, with the third line listing a surface brightness
survey needing to reach a 1-$\sigma$ limit of 1~K over 5~kHz channel under the
assumption of a $1'$ resolution. The final row lists the time necessary
to reach 1-$\sigma$ of 1~mJy across a 1~MHz bandwidth to a point source
at the centre of the field.
The expansion option for \tel\ 
offers a factor of almost 5 improvement over the strawman design.

\subsection{Comparison with other instruments}
The large field-of-view makes \tel\ an unprecedented synoptic radio
telescope, achieving survey speeds not available with any other telescope.
\tel\ will routinely image very large areas of the sky to sensitivities
only achievable with current instruments over very small areas.
\begin{table*}
\begin{center}
\caption{Sensitivity and survey speeds for \tel}
\begin{tabular}{lccc}
\hline & \vspace{-3mm} \\
\multicolumn{1}{c}{Parameter} & Strawman & Expansion\\
\hline & \vspace{-3mm} \\
Continuum survey speed (300 MHz, 100~$\mu$Jy) & 250 & 1150 & deg$^2$/hr\\
Line survey speed (100~kHz, 5~mJy) & 209 & 960 & deg$^2$/hr\\
Surface brightness survey speed (5~kHz, 1~K, $1'$) & 18 & 83 & deg$^2$/hr\\
Point source sensitivity (1~MHz, 1~mJy) & 1290 & 280 & sec\\
\hline & \vspace{-3mm} \\
\end{tabular}
\label{intro:speed}
\end{center}
\end{table*}

The survey speed of \tel\ exceeds that
of the Parkes 20-cm multibeam receiver, the Very Large Array (VLA) and
the Giant Metre Wave Telescope (GMRT) by more than an order of
magnitude for spectral line surveys in the GHz band.
In continuum, \tel\ can survey
the sky some 50 times faster than the NVSS survey carried out by
the VLA over a decade ago (Condon et al. 1998).

As an interferometer \tel\ provides low-frequency imaging
not otherwise available at the other major southern hemisphere interferometric
array, the Australia Telescope Compact Array (ATCA).
For single pointings it exceeds the ATCA sensitivity at 1400~MHz
and will also have better resolution and surface brightness sensitivity.
In terms of survey speed however, it gains by large factors for both line,
continuum and surface brightness sensitivity and will be comparable
to other planned facilities such as the Karoo Array Telescope (KAT) in
South Africa, the Allen Telescope Array (ATA; Deboer et al. 2004) in the 
USA and APERTIF in the Netherlands.

The survey speeds of \tel\ (with the strawman parameters),
the ATA and APERTIF are almost identical
although achieved in different ways. The ATA achieves its survey speed through
a large number of small dishes each with a single pixel feed.
APERTIF will have a focal plane array system on larger dishes.
The presence of telescopes with similar survey
speed in the northern hemisphere makes for excellent complementarity to \tel.

\subsection{Configuration of \tel}
A number of science projects (pulsar surveys, Galactic \hi, 
low surface brightness mapping) require a highly compact array configuration 
in order to increase the surface brightness survey speed
(see equations~\ref{eq:sb} and \ref{eq:ff}).
On the other hand science such as continuum and transients require
long baselines both to overcome the effects of confusion and to
obtain accurate positions for identification at other wavelengths.
In the middle is the extragalactic \HI\ survey which needs moderate
resolution to avoid over-resolving the sources.
With a total of only 30 dishes it is difficult to achieve all these
requirements simultaneously.

It is envisaged that three array configurations will be available
ranging from a very compact configuration (maximum baseline $\lesssim$400~m)
through a medium compact configuration (maximum baseline $\sim$2~km)
to an extended configuration (maximum baseline $\sim$8~km).
Configuration changes would occur by physically moving the antennas
and would happen only infrequently.

A design study is currently underway to determine exact antenna positions
for different configurations given the science case and the 
constraints of the local site topology and terrain.

\subsection{Location of \tel}
The central core of \tel\ will be located at the Murchison Radio 
Observatory in inland Western Australia, one of
the most radio-quiet locations on the Earth and one of
the sites selected by the international community as a potential
location for the SKA.
The approximate geographical coordinates of the site are longitude
116.5 east and latitude 26.7 south.
The southern latitude of \tel\
implies that the Galactic Centre will transit overhead and the
Magellanic Clouds will be prominent objects of study.
At least 30,000 square degrees of sky will be visible to \tel.
The choice of site
ensures that \tel\ will be largely free of the harmful effects of
radio interference currently plaguing the present generation of telescopes,
especially at frequencies around 1~GHz and below. Being able to
obtain a high continuous bandwidth at low frequencies is critical to
much of the science described in this document.

\subsection{\tel\ timeline}
In early 2008, the \tel\ test bed antenna will be installed at the site
of the Parkes radio telescope to allow testing of the focal plane
array and beamforming systems. Following this, the goal
is to have the first 6 antennas of \tel\ on-site in
Western Australia in early 2010. Over the subsequent two years
the remainder of the antennas would be deployed, with commissioning
of the final system expected to take place in 2012.

\section{Extragalactic \HI\ Science}
Understanding how galaxies form and evolve is one of the key
astrophysical problems for the 21st century. Since neutral hydrogen (\HI) is
a fundamental component in the formation of galaxies, being able to
observe and model this component is important in
achieving a deeper understanding of galaxy formation.
At any given time we expect that the amount of neutral hydrogen in a galaxy
will be determined by the competing rates at which \HI\ is depleted
(mostly by star formation) and replenished
(mostly by accretion of cold gas from its surroundings).
Understanding how the
abundance and distribution of \HI\ in the Universe evolves with redshift
therefore provides us with important insights into the physical
processes that drive the growth of galaxies and is a powerful test of
theoretical galaxy formation models (e.g. Baugh et al. 2004).

The cosmic \HI\ mass density $\Omega_{\rm HI}$ provides a convenient
measure of the abundance of \HI\ at a given epoch.
Estimates of $\Omega_{\rm HI}$ at high redshifts ($z \gtrsim 1.5$)
can be deduced from QSO absorption-line systems in general and damped
Lyman-$\alpha$ (DLA) systems in particular. DLA systems contain the
bulk of \HI\ at high redshifts and imply that $\Omega_{\rm HI}
\simeq 10^{-3}$ (e.g. P\'eroux et al. 2003, Prochaska et al. 2005,
Rao et al. 2006). At low-to-intermediate redshifts, however, the
only known way to measure $\Omega_{\rm HI}$ accurately is by means of
large-scale \HI\ 21-cm surveys.

This has been possible at $z \lesssim 0.04$ using the \HI\ Parkes All-Sky
Survey HIPASS (Koribalski et al. 2004; Meyer et al. 2004), which mapped the
distribution of \HI\ in the nearby Universe that is observable from
the Parkes radio telescope.  HIPASS data have allowed accurate measurement
of the local \HI\ mass function (HIMF) and $\Omega_{\rm HI}$
of galaxies (Zwaan et al. 2003, 2005).
However, few measurements of the HIMF and $\Omega_{\rm HI}$ using \HI\
21-cm emission have been possible over the redshift range
$0.04 \lesssim z \lesssim 1$ because of insufficient sensitivity of
current-generation radio telescopes (though see e.g. Verheijen et al. 2007).
In the simulations that follow, it is therefore assumed that the HIMF does
not evolve with redshift. This assumption is conservative and any evolution
in the HIMF will likely result in the detection of a greater
number of \HI\ galaxies.
\begin{figure}[h]
\centerline{\psfig{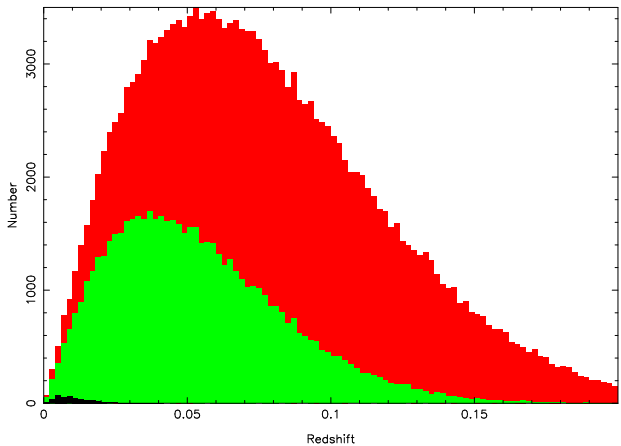}}
\caption{Number of galaxies above $5\sigma$ as a function of redshift
bin for a shallow
\tel\ \HI\ survey lasting a year, covering the southern hemisphere, compared
with HIPASS HICAT (Meyer et al. 2004; black histogram). The
simulation assumes the Zwaan et al. (2005) HIMF with no evolution, and a
bivariate relation between \HI\ mass and velocity width.
The simulations further assume
a compact \tel\ configuration -- the number of detections
sharply reduces for angular resolutions below $1'$.
The number of predicted detections is $\sim1.8\times10^6$ and
$\sim0.6\times10^6$ for the 
expansion (red histogram) and strawman (green histogram) options.}
\label{fig:hishallow}
\end{figure}

Widefield \HI\ surveys using the next generation radio telescopes 
such as \tel\ and
ultimately the SKA will allow unprecedented insights into the
evolution of the abundance and distribution of \HI\ with
cosmic time, and its consequences for the cosmic star formation, the
structure of galaxies and the Intergalactic Medium.
\tel\ excels as a survey telescope as it will be able to spend long
periods of time integrating on large areas of sky, resulting in the
detection of large numbers of galaxies. Two compelling \HI\ surveys are:

\begin{itemize}
\item A shallow hemispheric \HI\ survey lasting a year. This would result in
the detection of over 600,000 galaxies, or two orders of magnitude greater
than HIPASS. The typical survey depth would be $z\sim 0.05$ with massive
galaxies detected out to $z\sim 0.15$ (see Figure~\ref{fig:hishallow}).
\item A deep survey covering a single pointing, also lasting
a year. Although this would only detect 100,000 galaxies in the example
of a non-evolving HIMF shown in Figure~\ref{fig:hideep}, the
typical depth would be $z\sim 0.2$ with massive galaxies detected out to
$z\sim 0.7$.
\begin{figure}
\centerline{\psfig{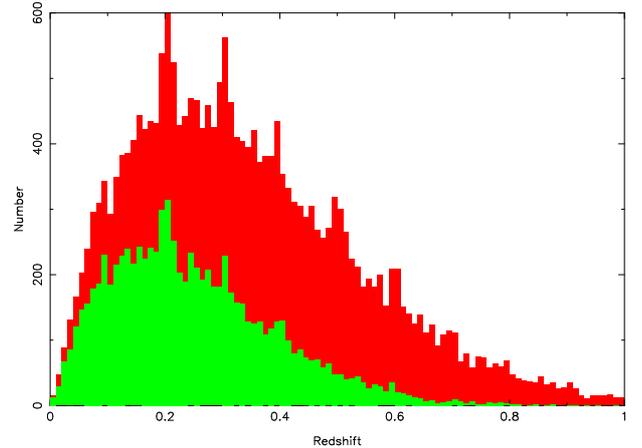}}
\caption{Number of galaxies above $5\sigma$ as a function of redshift
bin for a deep \tel\ \HI\ survey lasting a year, covering a single pointing. The
simulation is similar to that for the all-sky survey and assumes a
non-evolving HIMF. The number of predicted detections is
$\sim 226,000$ for the expansion option (red histogram) and $\sim 99,000$ for
the strawman option (green histogram). There is less dependence on angular
resolution than for the shallow all-sky survey. Standard cosmology is
assumed (flat Universe, H$_o=75$ km s$^{-1}$ Mpc$^{-1}$, $\Omega_m=0.3$,
$\Omega_{\Lambda}=0.7$).}
\label{fig:hideep}
\end{figure}
\end{itemize}

The results from this survey will
provide powerful tests of theoretical galaxy formation models and
improve our understanding of the physical processes that shaped the
galaxy population over the last $\sim$7~Gyr.

\tel\ will provide useful cosmological measurements,
but the detection of the so-called
baryonic acoustic oscillations will probably only be
possible on small spatial scales with the
all-sky survey. This will preclude the investigation of dark energy
studies at higher redshift in a manner competitive
with future and ongoing optical surveys such as `WiggleZ'
(Glazebrook et al. 2007).
The leading sources of systematic error for baryon
oscillation surveys are the poorly-modelled effects of redshift-space
distortions, halo bias and the non-linear growth of structure.  These
processes all modify the underlying linear power spectrum which
encodes the baryon oscillation signature, hampering our cosmological
measurements.  \tel\ can make inroads into this problem and 
pave the way for future SKA surveys by 
precisely measuring the clustering
properties of \HI\ galaxies over a range of redshifts
and through measuring the HIMF with
unprecedented accuracy, including how this function varies with
redshift (to $z \sim 0.7$) and environment. 

\section{Continuum Science}
\label{sec:cont}
Understanding the formation and evolution of galaxies and active galactic
nuclei (AGN) as a function of cosmic time is a key science driver
for next-generation telescopes at all wavebands. Today's
instruments already give profound insights into the galaxy population at
high redshifts, and a number of current surveys
are asking questions such as: When did most stars form? How do AGN
influence star formation? What is the spatial distribution of evolved
galaxies, starbursts, and AGN at $0.5 < z < 3$? Are massive black holes a
cause or a consequence of galaxy formation?

However, most of these surveys are primarily at optical and 
infra-red wavelengths, and can be significantly misled by dust extinction.
A survey with \tel\ will be able to determine how
galaxies formed and evolved through cosmic time, by penetrating the
heavy dust extinction which is found in AGN at all redshifts, and
studying the star formation activity and AGN buried within.
An \tel\ survey, with baselines out to $\sim$8~km ($5''$ resolution)
can survey the entire southern sky to
a flux limit of $\sim50~\mu$Jy\,beam$^{-1}$ in one year, a limit which is
currently being reached only in tiny parts of the sky (e.g.
in the HDF-N and HDF-S deep fields).
Such a survey is likely to have the enormous impact that the NVSS
(Condon et al. 1998) has had over the past decade,
but at a factor of 50 better in sensitivity and 9 in angular resolution.

With a $5\sigma$ flux density limit of $50\,\mu$Jy\,beam$^{-1}$,
typical starburst
galaxies, with star formation rates (SFRs) around $100\,$M$_{\odot}$yr$^{-1}$,
will be detectable to $z\sim2$. The most extreme starbursts, with SFRs of a
few $1000\,$M$_{\odot}$yr$^{-1}$, will be visible well into the epoch of
reionisation ($z \sim 6-10$) if indeed they existed
then. Ordinary disk galaxies like the Milky Way, with SFRs of only a few
M$_{\odot}$yr$^{-1}$, will be visible to $z \sim 0.3$.
A confusion limited sky survey with the 8~km
baseline configuration, obtained in addition to a compact configuration
for the \HI\ survey,
will make the star forming galaxy population accessible as never before. It
will allow unprecedented exploration into how star formation in galaxies
evolves with time, and how it depends on galaxy mass, galaxy environment,
past star formation history, interaction/merger history, and more.

Another important factor to consider is distinguishing
the star forming population from the AGN population. At flux densities
$\lesssim1\,$mJy starburst galaxies start to become a major component of the
1.4\,GHz source counts, dominating below $0.5\,$mJy or so (Hopkins et al. 2000;
Jackson 2005). Recent measurements, though, suggest that there may still be a
significant proportion of low-luminosity AGN (Simpson et al. 2006) even at these
levels. By $0.1\,$mJy, existing AGN evolutionary models imply that starburst
galaxies should be dominant, although normal star forming galaxies are not
expected to dominate the counts until levels below $\sim1\,\mu$Jy
are reached (Windhorst et al. 1999; Hopkins et al. 2000).

There is an inevitable degeneracy between luminosity and redshift
inherent in flux limited samples of classical (active) radio galaxies.
This arises because only the most powerful sources are detected at the
largest redshifts. In order to explore the physics of radio galaxies, and
especially their evolution, large samples of radio galaxies of
comparable luminosity across cosmic time are required. 
At present, we are still unable to say much about the global population of 
radio galaxies at $z\gtrsim0.5$ because the faint source
counts are so limited.  

The results of the proposed \tel\ confusion limited survey, 
coupled with optical spectroscopic data from e.g. PanStarrs and the 
LSST, will generate samples from which luminosity functions can be 
determined for FRI and FRII galaxies as a function of $z$, as well as 
for the high-redshift radio galaxy population that are strikingly 
different to the canonical FRI and FRII morphologies we are used to 
observing at $z\lesssim1$ (van Breugel et al. 1999). Note that \tel\ will be 
able to detect FRIs to z$\sim$4 based on their luminosity alone, but only
out to z$\sim$1.7 based also on their morphology.

The resulting science should not be underestimated: radio
galaxies represent the most massive galaxies at any redshift up 
to $z\sim5.2$ (Rocca-Volmerange et al. 2004) and
therefore represent our best chance to understand the formation and evolution of
massive galaxies and their central supermassive black holes. In addition,
they are likely also to provide the dominant source of energy in
the universe, and possibly even provide the dominant source of magnetic fields.

It is worth emphasising that at $5''$ resolution, incompleteness
becomes significant as emission extended on scales larger than this is
resolved (see the discussion regarding the choice of angular resolution
for the NVSS in Condon et al. 1998). However, 
this issue is likely to be less of a problem for \tel\
than for NVSS for two reasons. First,  at the extremely faint
flux densities of the proposed \tel\ survey, the source counts are
dominated by distant star-forming galaxies with $\sim1''$ angular sizes, rather
than the larger and more powerful radio galaxies which dominate
the source counts in NVSS. Second, a continuum all-sky survey with short
baselines will be available as a by-product of the proposed \HI\ sky survey.
While this compact continuum survey will be confusion-limited, the data
can be sensibly combined with the longer-baseline $5''$ resolution southern
sky survey to create a complete {\textit{and}} confusion-limited
high-resolution sky survey. Furthermore, the low surface brightness
survey  will be sensitive
to the diffuse emission from, for example, dying radio galaxies
(Murgia et al. 2005) cluster haloes, relics and ghosts
(e.g. Feretti 2000, En{\ss}lin 1999, En{\ss}lin \& Gopal-Krishna 2001),
`fat double' FRI tails and plumes (Subrahmanyan et al. 2006), and
perhaps even the Thomson scattered electrons off 
dark matter (Geller et al. 2000).

\tel\ may answer directly the question of whether dark energy 
exists at $z\sim1$, and if so on what scale.  A unique independent
test of dark energy is to observe a correlation between CMB fluctuations and 
low-redshift large-scale structure (Crittenden \& Turok 1996), the so-called
late-time Integrated Sachs-Wolfe (ISW) effect.
Such a correlation would only be seen if the CMB photons have been 
redshifted by the low-redshift structure as it collapses, and only 
occurs if the current expansion of the Universe is {\textsc{not}} 
matter dominated. Cross-correlation of the
NVSS radio galaxies with the CMB anisotropies gave a tentative detection of the
ISW effect, but at significance of only $2\sigma$ (Boughn \& Crittenden 2004).
A more recent analysis (Pietrobon et al. 2006) used WMAP3 data and powerful
statistical techniques to exclude the null hypothesis at 99.7\%.
However an extremely deep survey with \tel\ can provide a
far more stringent examination; this test has a much lower confusion limit than
the traditional limit for source counts because we are looking for a
statistical result on size scales of order the scales of the CMB anisotropies.
Moreover, the deep \tel\ survey may also provide the ability to 
track the effect as a function of redshift.

\section{Polarization Science}
One of the five key-science drivers for the SKA is
to understand the origin and evolution of cosmic magnetism.
The history of the Universe from the Big Bang to the present day is
essentially a history of the assembly of gas into galaxies, and the
transformation of this gas into stars and planets.  While gravitation
initiates and sustains these processes, it is magnetism which breaks
gravity's symmetry and which provides the pathway to the non-thermal
Universe.  By enabling processes such as anisotropic pressure support,
particle acceleration, and jet collimation, magnetism regulates the
feedback that is vital for returning matter to the interstellar and
intergalactic medium (Boulares \& Cox 1990, Zweibel \& Heiles 1997).

Prime observational approaches to this key science will include a deep
survey of polarized emission from extragalactic sources (EGS) (Beck
\& Gaensler 2004), and of the diffuse polarized radiation from the
Galaxy. These surveys will be extremely important in the study of
Galactic magnetic fields, both within the Milky Way and in external
galaxies. A key experiment with \tel\ will be to
image the entire southern sky at 1.4~GHz to a 1-$\sigma$ sensitivity of
10~$\mu$Jy in full-Stokes continuum emission as described in 
section~\ref{sec:cont}. The polarization products from this survey
would provide

\begin{itemize}
\item a deep census of the polarization properties of galaxies as a
function of redshift (secured through complementary \HI\ or optical surveys),
\item a dense grid of Faraday rotation measures to over
500,000 background radio sources
\item all-sky Faraday Rotation image of the 3-dimensional structure of the 
diffuse magneto-ionic medium of the Galaxy.
\end{itemize}

The fractional polarization, $\Pi$, of EGS is typically only a few 
percent, so the signal to noise ratio in
polarized flux is always a factor $\sim 50$ below the signal to noise
ratio in total intensity. 
This has limited investigations of the polarization
properties of EGS to the bright ($\gtrsim 80$~mJy) radio source population,
dominated by powerful, radio
loud AGN. There are some indications from these bright sources that at
least fainter steep-spectrum radio sources are more highly polarized
(Mesa et~al.\ 2002; Tucci et~al.\ 2004), up to a median $\Pi$ of $\sim1.8\%$
in the flux density range 100 -- 200 mJy and even higher for sources
below 100~mJy (Taylor et al. 2007).

\tel\ will revolutionize this field by providing larger samples of
polarized sources, to much fainter flux densities than currently
possible. At a sensitivity level of 10 $\mu$Jy, \tel\ will begin to
reveal the polarization of radio sources as faint as a few mJy, where
star forming galaxies begin to contribute measurably to the total
radio source counts (Windhorst 2003).
Models of the integrated polarized emission from disk galaxies, based on
magnetic field strengths and geometries, in
conjunction with analysis of the integrated polarization of nearby disk
galaxies, will  provide a theoretical basis for analysis of the magnetic
properties of the large sample of distant star-forming galaxies that will be
observed by \tel\ (Stil et al. in prep).  Idealized models
suggest that the $\Pi$ distribution of a
large sample of randomly oriented disks has a minimum at $\Pi =
0$. The presence of such a minimum in the distribution depends on the
importance of Faraday depolarization and the importance of poloidal
magnetic fields in the halo of these galaxies. We can measure this
minimum by including the lowest-order antisymmetric deviations from a
Gaussian in our Monte-Carlo analysis of a sufficiently large
number of galaxies (Taylor et al. 2007).
The projected large frequency coverage of \tel\
will allow us to distinguish between highly frequency-dependent
Faraday depolarization, and magnetic field structure that should not
change with observing frequency. As we consider integrated quantities,
our analysis will not be affected by variation of resolution with
frequency.

The Milky Way and many other nearby spiral galaxies all show
well-organized, large-scale magnetic fields (Beck et al. 1996), for
which the dynamo mechanism is the favoured explanation
(Ruzmaikin et al. 1988).
However, dynamos are not yet well understood and still face theoretical
difficulties, especially in light of recent results which show that field
amplification in galaxies can be extremely rapid (Gaensler et al. 2005).
Our own Milky Way is an excellent test-bed to better study the underlying
physical processes, because a huge ensemble of background Faraday
rotation measures (RMs) can be used to probe its three-dimensional
magnetic field structure.  RMs from both pulsars and EGS
should be used in such studies. Until recently (Brown et al. 2003, 2007)
there had been comparatively few EGS RMs available, while studies 
utilising pulsar RMs
alone have been limited both by the comparatively sparse sampling
of pulsars on the sky and by uncertainties in pulsar distances. 
As a result, mapping the three-dimensional magnetic
field distribution has been difficult, especially in complicated regions.
\begin{figure}[ht]
\centerline{\psfig{file=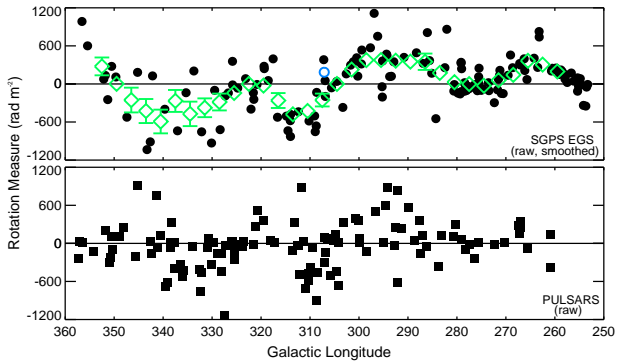,width=0.5\textwidth}}
\caption{RM vs Galactic longitude for EGS and pulsar RMs in the Southern 
Galactic Plane Survey (SGPS; Brown et al. 2007).
Top panel: Circles represent individual RMs of EGS while open diamonds 
represent these data boxcar-averaged (smoothed) over 9$^\circ$ in 
longitude with a step size of 3$^\circ$.  Where symbol size permits,
the error bars are the standard error of the mean.
The single open circle at $l=307^\circ$ represents the only EGS
in this region prior to the SGPS.
Bottom panel: squares represent individual pulsars with known
RMs in the SGPS region. The oscillations seen in the data reveal
the orientation of the large-scale magnetic field in the spiral arms and
interarm regions.}
\label{fig_brown}
\end{figure}

Broadly speaking, the Milky Way's magnetic field has two components:
a large-scale, coherent, magnetic field, tied to the overall structure
of the spiral arms, disk and halo, and a small-scale, fluctuating field
which traces supernova remnants (SNRs), \HII\ regions,
and diffuse turbulence in the ionised ISM (Ruzmaikin et al. 1988).
One method of separating the smooth and fluctuating components of the magnetic
field is to average the RM data from EGS into intervals of
$\sim20$~deg$^2$ as shown in Figure~\ref{fig_brown}.
Such an analysis begins to beautifully reveal the
overall structure associated with field reversals and spiral arms.
However, \tel\ can greatly improve
current RM yields, providing a source density over the entire sky of
$\sim20$~RMs per deg$^2$.  With such a data-set, we can successfully
distinguish between the behaviours of the smooth and fluctuating
components of the field (at a signal-to-noise ratio of $\sim4-5$) down
to a resolution of $\sim1$~deg$^2$. One square degree turns out not to be
some arbitrary scale, but represents an approximate dividing line between
the sizes of individual SNRs and H\,{\sc ii}\ regions, and the larger
scale structure of spiral arms, super-bubbles and fountains. Polarimetry
with \tel\ will thus provide our first complete view of the magnetic
geometry of the Milky Way, on scales ranging from sub-parsec turbulence
up to the global structure of the disc and spiral arms.

\section{Galactic and Magellanic Science}
The renaissance of observational studies of the Milky Way and Magellanic
System over the past decade has raised new and profound
questions about the evolution of the interstellar medium (ISM).
The community has transitioned from studying small-scale aspects
of the ISM to a more comprehensive approach, which seeks to combine
information about a variety
of ISM phases with information about magnetic fields.  With \tel\ we can make
significant and unique inroads into understanding the evolution of the
ISM and through that the evolution of the Milky Way.  These are crucial
steps along the path to understanding the evolution of galaxies.

Galaxy evolution is one of the great puzzles in current astrophysics,
incorporating how galaxies assemble and evolve from the beginning of
the universe to the present day.  Unfortunately, our knowledge of the
fundamental processes of galaxy evolution is blocked because we do not
understand the evolutionary cycle of interstellar matter in our own
Galaxy. We know that the life cycle of the Milky Way
involves a constant process of stars ejecting
matter and energy into the interstellar mix, from which new stars then
condense.  Somehow matter makes the transition from hot, ionised
stellar by-products to become the cold molecular clouds from which
stars are formed. The circulation of matter between the Galaxy's disk
and its surrounding halo further complicates this cycle. ISM studies
in our Galaxy probe the evolutionary cycle with sensitivity and
resolution unattainable in external galaxies; it is only in the Milky
Way that we can observe the evolution of the interstellar medium on
scales ranging from sub-parsec to kiloparsecs. Furthermore, the
evolution of the Galaxy is controlled by the Galactic and local
magnetic fields, yet these components are largely unknown.  The Milky
Way is thus an ideal laboratory for studying galaxy evolution.

In recent years there has been a renaissance in Galactic
ISM surveys.  This has largely been driven by advances with
the Canadian and Southern Galactic Plane Surveys, which used a
combination of single antenna and aperture synthesis telescopes to
map the Galactic plane in \HI\ and polarized
continuum emission at an unprecedented resolution of $\sim1'$.
These results can be combined with a number of Galactic Plane surveys
at comparable or better resolution in $^{12}$CO (Clemens et al. 1988), 
$^{13}$CO (Jackson et al. 2006) and the Spitzer GLIMPSE
survey of infrared emission. These surveys have 
revealed structure on all size scales, a
remarkable agreement between ISM phases, and a wide variety of new
polarization structures.  

The Milky Way and Magellanic System, because of their very large sky
coverage, can only be observed in survey mode. 
An all-sky Galactic \HI\ survey with \tel\ would build
on current surveys to provide a full census of all \HI\ associated with 
the Milky Way and the Magellanic System at arcmin resolution. 
We therefore propose a 1~year survey to a brightness temperature sensitivity 
limit of $\sim100$~mK over a $\sim2'$ synthesised beam using the 
very compact configuration.
An additional consideration for this survey is the need to
include single dish data for sensitivity to structure on the largest scales.
Fortunately the Galactic All-Sky Survey already exists with comparable
sensitivity and spectral resolution to the proposed survey and can be
combined with the \tel\ survey.
The survey would
provide essential information about the physical and thermal structure
of high velocity clouds and their interaction with the halo;
the role and origin of halo cloudlets and the physical structure of 
the Magellanic Stream and Leading Arm.

One of the key science drivers for the SKA is
understanding the magnetic universe.  For the Milky Way, rotation
measures of extragalactic point sources
will yield information about the line of sight averaged
magnetic field weighted to regions of ionised gas.  To fully
understand the Galactic field and its effect on the evolution of the
Galaxy we propose two surveys, (a) an all-sky survey of the polarization 
state of the diffuse emission to provide data for detailed study of
the magneto-ionic medium (MIM), and (b) a survey of Zeeman splitting
of the \HI\ line to provide {\em in situ} magnetic field measurements.

The appearance of the polarized sky is dominated by Faraday rotation.
Existing data permit this conclusion about the MIM, but are entirely
inadequate for unravelling its detailed physical characteristics.  The
surveys that exist are of very poor angular resolution or provide only
interferometric data without complementary single-antenna data, and
hence miss important information on broad structure.
Wideband, multi-channel imaging of the diffuse polarized emission
promises to provide significant new data on the MIM through the 
technique of Faraday rotation synthesis (Brentjens \& de Bruyn 2006). 
Such a survey with \tel\ will allow 
identification of RM features with known objects (stars and various 
kinds of stellar-wind bubbles, SNRs, low-density envelopes of \hii\
regions, ionized skins on the outside of molecular clouds), and
promises to give information on the turbulent structure of the
MIM. The latter outcome will shed light on the role of magnetic fields
in the turbulent cascade of energy from large to small scales. Such a
survey will also give measures of field strength in many regions. Most
polarization surveys have focused on the Galactic disk, and the \tel\
survey will give important new information on the disk-halo transition
and on conditions high above the Galactic plane.

Zeeman splitting of the \HI\ line can provide {\em in situ} magnetic field 
measurements throughout the Galactic disk (e.g. Heiles \& Troland 2003, 2005).
Ideally, one would like to measure the Zeeman splitting of the \HI\ emitting
gas at the tangent points for all longitudes between $-90^{\circ}$
and $+90^{\circ}$.
A small, but significant latitude coverage of $\sim
5^{\circ}$ is needed to ensure that the measurements reflect the
global Milky Way magnetic field, rather than being subject to
individual  clouds.  We therefore propose an \tel\ survey of the
area $270^{\circ} \leq l \leq 45^{\circ}$, $|b|\leq 2.5^{\circ}$:
$\sim700~{\rm deg^2}$. For a typical magnetic field strength of
$\sim 5~{\rm \mu G}$ and a velocity width of 5 \kms, we expect the amplitude
of the Stokes V spectrum to be 70~mK.  We require that the derivative
of the line be very well sampled in velocity space, so for a typical
derivative line width of 5 \kms, we require $\Delta v \approx 0.25$
\kms or $\Delta \nu \approx 1$ kHz.  This survey requires a 1-$\sigma$
sensitivity limit of $\sim 20$ mK in each polarization.  In order to
reach these low surface brightness limits we once again require that
the telescope have a high filling factor.  Because resolution is not a
stringent demand on this project, an ultra-compact configuration with
most of the collecting area inside a maximum baseline of $\sim 200$ m
would be preferred.  Measuring \HI\ Zeeman in emission requires both
low sensitivity limits and good polarization properties of the
telescope.  With such a compact configuration \tel\ would be uniquely
positioned to measure the magnetic field along the tangent point and
contribute directly to constraining models for the Milky Way magnetic
field.

\section{VLBI Science}
\begin{figure}[h]
\centerline{\psfig{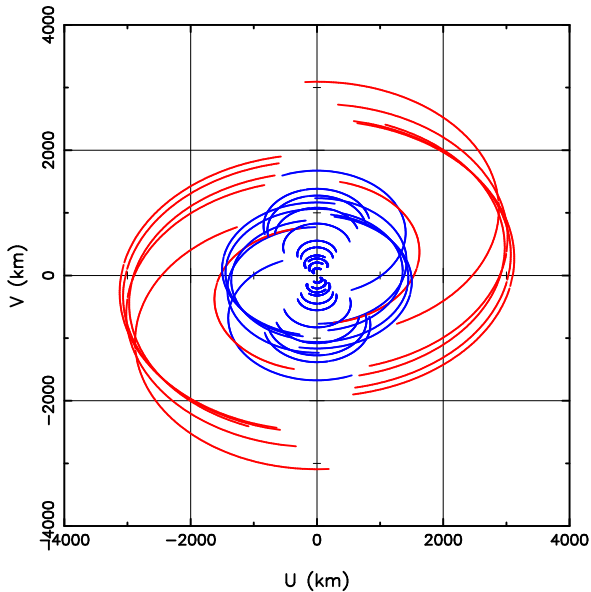}}
\caption{The (u,v) coverage of the current Australian VLBI array at 
1.6 GHz (blue curves) and with the inclusion of ASKAP (blue and red
curves combined). The improvements in the east-west component of the baseline 
distribution and in the extension of the maximum
baseline length are significant.}
\label{vlbi_fig1}
\end{figure}

A number of scientific programs
present themselves when considering the use of \tel\
as part of the Australian Long Baseline Array (LBA) and the
global VLBI array.  Better angular resolution at 1.4 GHz, better
sensitivity, and better (u,v) coverage (see Fig~\ref{vlbi_fig1})
will aid standard VLBI observations
of AGN, pulsars, and OH masers.  An innovative additional capability
for \tel\ is multibeaming.  If this can be harnessed for VLBI in the form of
multiple phased array beams, a number of wide-field survey observations become
feasible.

\begin{figure}[h]
\centerline{\psfig{file=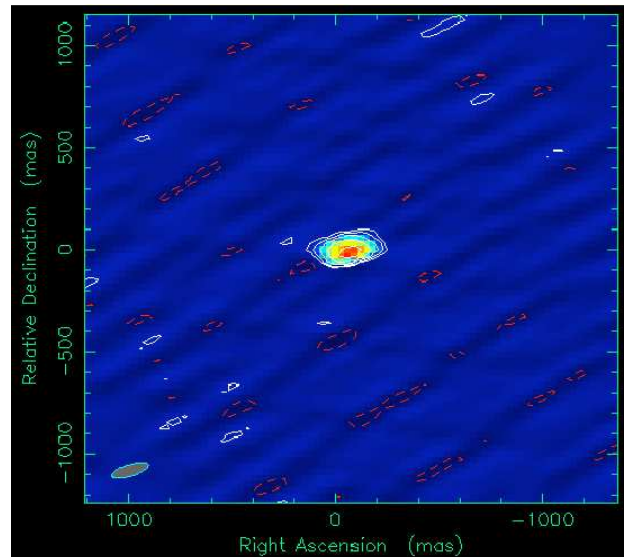,angle=0,width=0.5\textwidth}}
\caption{The PAMHELA image of Circinus X-1, the first Australian e-VLBI image
of a transient radio source associated with a Galactic X-ray binary object
(Phillips et al. 2007).}
\label{vlbi_fig2}
\end{figure}

In time, \tel\ should also become a part of the recently
developed Australian e-VLBI network, currently called PAMHELA
(see Fig~\ref{vlbi_fig2}).
An interesting possibility is to not use \tel\ as part of the LBA, but use
it as a source of trigger information for radio transients, as part of
\tel\ survey work.  These triggers could be transmitted to PAMHELA, and the
candidate sources targeted at high angular resolution in rapid follow-up
observations.  The combination of \tel\ and
PAMHELA would be unique and powerful.  PAMHELA would add great scientific
value to the low resolution detection of transients by \tel.

In the 1.4 GHz band, a number of VLBI science applications are
possible, using \tel\ as an additional element in the LBA.
Given the phased array field of view of approximately $5''$, the
following observations could be usefully made:
\begin{itemize}
\item Active galactic nuclei (AGN) only require a narrow field of view.  The
addition of long, sensitive east-west baselines is critical for performing high
dynamic range imaging of the often complex mas-scale jet radio sources.
Further, use of \tel\ along with RadioAstron observations will be possible,
examining AGN on the longest possible baselines to investigate brightness
temperature limits in AGN.
\item  Pulsar proper motion and parallax observations benefit from good
sensitivity on long baselines in order to do the best possible astrometry
(e.g. Brisken et al. 2003).
The relatively low frequency of \tel\ is optimal for pulsars. The combination
of longer baselines and better sensitivity would allow the LBA to target a much 
wider range of pulsars for astrometry than is presently feasible. 
\item OH Maser observations benefit from good sensitivity on long baselines,
important to resolve individual maser spots. Different maser species 
typically trace different parts of the star forming regions.  
Combining high-resolution 
OH imaging with observations of methanol and water masers, and the 
high velocity resolution dynamical information it is possible to 
image and interpret the three-dimensional distribution of masers in great 
detail, and in some cases to obtain proper motions and accurate 
distance estimates (e.g. Boboltz 2005).
\item The detailed study of rare compact objects, such as supernovae seen at 
both optical and radio wavelengths (e.g. Bietenholz 2005)
\end{itemize}

A significant part of the \tel\ science case involves the rapid
survey capability of the instrument and its ability to detect rapid
transients in continuum emission.  The types of transient that \tel\ will
likely detect include Galactic X-ray binaries,
gamma-ray bursts (GRB), flaring stellar systems, and possibly
supernovae in nearby galaxies. A fuller description of transient searches
with ASKAP follows in Section~\ref{sec:trans}.

Both the new types of transients that ASKAP will likely discover, and
the previously known types of transients are likely to involve objects
of small angular size and varying structure such as
X-ray binaries, GRBs, and supernovae.  For all these objects,
especially any newly discovered types of transients, therefore, VLBI
observations will prove useful to elucidate the source structure and
its evolution.  Such VLBI observations will be particularly useful if
the source is imaged while in an active state, and so the combination
of ASKAP transient event detection with rapid-response PAMHELA
observations should pay particularly high science dividends.

\section{Pulsar Science}
Almost 2000 radio pulsars have been discovered since the
initial detection in Cambridge (Hewish et al. 1968).
Pulsars have been used as tools to address
some of the most fundamental questions in basic physics (allowing
precision tests of general relativity, investigations into the
equation of state of ultradense matter and the behaviour of matter and
radiation in the highest magnetic fields known in the universe)
and astrophysics (binary evolution, binary dynamics, the
interstellar medium, globular cluster physics, supernova remnant
astrophysics, the physics of relativistic winds and precision astrometry).
Pulsars are also intrinsically interesting, being the result of core collapse
supernovae and astonishing converters of mechanical energy of
rotation into electromagnetic radiation, particles and magnetic
fields.  The study of pulsars themselves is important for constraining
the overall population's properties and hence their origin, as well as
understanding the mysterious pulsar emission mechanism.

Historically it has been common to carry out pulsar research using
large single dish instruments. \tel\ will provide the transition from using
large diameter single dishes to using large numbers of small antennas with wide
field-of-view (FoV) capability as is likely in the final mid to high frequency
SKA design. However, 
pulsar observations, especially searches, with \tel\ will present
myriad computational challenges, some of which can be
mitigated through specification choices.  In particular, any
large-scale survey with good sensitivity will be severely
computationally limited, and the need to process every pixel
independently ensures that the computational load becomes
larger as the square of the
maximum baseline length.  It is therefore necessary that as many short
baselines as possible be present in the configuration of \tel. For
pulsar timing, there is no such requirement; any configuration is
adequate.

We suggest a survey with \tel\ which occupies $\sim$100~days of
observing time and covers the 30,000 square degrees of visible sky.
Each 30 square degree single pointing therefore is observed for
$\sim$150~min. Such a survey, with the strawman \tel\ parameters,
would equal the sensitivity of the Parkes Multi-beam Pulsar Survey 
(but over a much larger area of sky) and
would be 10 times more sensitive than the previous Parkes all-sky
survey.  Data recording requirements would force the
survey to sample at a rate of only 5 or 10~ms; this would necessarily
and unfortunately preclude sensitivity to millisecond pulsars. 

Simulations (Lorimer; private communication) show that such a survey
would detect 1600 pulsars with periods $\gtrsim40$~ms, about
half of which would be new discoveries. Many of the new detections
would be low-luminosity objects; a large sample of low-luminosity
pulsars is important in defining pulsar birthrates and
evolution. Furthermore, even though this proposed survey would not be
sensitive to millisecond pulsars, many exotic objects have longer
periods including the original binary pulsar PSR~B1913+16 (Hulse \&
Taylor 1975), pulsars with high-mass companions and relativistic
systems such as PSR~J1141$-$6545 (Kaspi et al. 2000).  It is likely
that any pulsar companion to a black hole in the field of the galaxy
would be slowly spinning and
the survey described here might find such an object.  Finally this
survey would be sensitive to transients with pulse widths greater than
10\,ms or so.
The science case for detecting new pulsars is
wide-ranging and includes:
\begin{itemize}
\item The ability to improve models of the pulsar population and hence
determine birth rates and the Galactic distribution
(e.g. Vranesevic et al. 2004.)
\item The discovery of unique objects, particularly highly
relativistic binary systems including pulsar-black hole binary
systems. Many unexpected discoveries will be made in future surveys.
\item Mapping the electron density distribution and
magnetic field of the Galaxy 
by combining rotation and dispersion measures with independent distance
estimates (e.g. Cordes \& Lazio 2002, Han et al. 2006.).
\item Understanding the pulse emission mechanism.  As the number of
known pulsars increases so does the variety in pulse profiles,
polarisation and fluctuation properties.  A more complete
understanding of the pulse emission mechanism will only occur from the
careful analysis of these properties for a large sample of pulsars at
multiple observing frequencies (e.g. Karastergiou \& Johnston 2006).
\item Timing of these newly discovered pulsars will be more efficient
with \tel\ than existing telescopes because the large FoV makes
it possible to time multiple pulsars simultaneously.
\end{itemize}

One of the most exciting modern-day applications of pulsar
timing is to combine data from multiple millisecond pulsars to form a global
timing array (Foster \& Backer 1990; Hobbs 2005). The aims of such a
project are many-fold with the main goal of making a detection of the
gravitational wave (GW) background at nano-Hertz frequencies.  In
brief, GW backgrounds are predicted to occur due to cosmological
(e.g. due to inflation, cosmic strings or phase transitions), or
astrophysical (e.g. due to coalescing massive black hole binary
systems that result from the mergers of their host galaxies) processes
(e.g.  Maggiore 2000).  The background is detected by looking for
correlations between the timing residuals of pulsars that have a wide
range of angular separations (Jenet et al. 2005).
\tel\ will enhance the capabilities of a global pulsar timing array
and demonstrate the possibility of high-precision pulsar timing
on an SKA-type instrument.

High precision millisecond pulsar timing will also continue to
undertake sensitive observations of relativistic effects in
double-neutron-star systems which lead directly to stringent tests of
relativistic gravity. \tel\ will have uniquely sensitive access to
many of the binary pulsar systems discovered in the Parkes multibeam
surveys.  Masses measured by \tel\ will add to the currently poor
statistics of masses in pulsar--white-dwarf systems and finally allow
a realistic investigation of mass dependencies on orbital period,
companion type and evolutionary history.

\section{The Transient Radio Sky}
\label{sec:trans}
It has been the general trend in radio astronomy to move from
dipole-like antennas towards parabolic dishes which have much
larger forward gain at the expense of a much smaller field-of-view.
Consequently, this
severely limits the possibility of detecting (random) transient events
and the transient sky in the radio is only poorly characterised.
At the same time, many classes of objects are known to be variable
radio sources including the Sun, the planets, cool stars, stellar
binary systems, pulsars, supernovae (SN), gamma-ray bursts (GRBs)
and active galactic nuclei (AGN).

The key to a successful transient instrument is to have high
sensitivity, large field-of-view, good dynamic range and high resolution.
\tel\ fulfills these criteria with its ability to achieve sub mJy
sensitivity across the entire sky in a single day observing.
Nearly all transients arise from point source objects; high resolution
is ideal for obtaining accurate positions necessary for follow-up
at other wavebands. It is important also that a wide range of timescales
from seconds to months are covered by the transient detector. This implies
a careful search strategy for uncovering rare objects.
\tel\ will most likely suffer from a surfeit of transient
and variable sources.
This will pose challenges both for imaging and for determining which
sources are most interesting and worthy of follow-up observations
on other facilities.

The most interesting transient sources detected with \tel\ will
undoubtedly be objects which we currently know nothing about. One of
the advantages of an all-sky survey is that it is tailor-made
for detecting the unknown. It has become clear recently that the
radio sky contains many transient objects, the identification of
which remain mysterious. For example, Hyman et al. (2005) discovered
a bursting transient towards the Galactic Centre which lasted for only
a few minutes but had a flux density in excess of 2~Jy. The bursts repeat
at irregular intervals and the identification of this source remains
unclear.  Bower et al. (2007) examined archival VLA data
of the same field spanning 22 years with observations approximately
once per week. They detected 10 transient sources, at least six of which
have no optical counterparts or quiescent radio emission. Bower et al. (2007)
consider the classes of known transients and conclude that their sample
is unlikely to be drawn from the known population.
They estimate that at \tel\ sensitivity, approximately 1 transient
source per square degree, with a duration of order a week, will be
present at any given time.

As an example of transient sources, consider gamma-ray bursts (GRBs)
which, in the gamma-ray band, occur approximately
twice per day and are visible out to $z\gtrsim 6$.
GRBs are almost certainly beamed in gamma-rays; this implies that we
only detect some fraction of the total population. However, it may
be possible to detect so-called orphan GRBs without
the initial gamma-ray trigger by looking for variable radio sources.
Levison et al. (2002) and Totani \& Panaitescu (2002)
show that several hundred radio afterglows of GRBs should
be present in the sky at any one time above a level of 1~mJy.
\tel\ could survey the sky to this level every day, providing a 
definitive test of these ideas.

Monitoring the variability in AGN is made possible by the wide FoV of \tel.
A survey with \tel\ could cover the sky every day to a $5\sigma$
limit of 2\,mJy, which would detect variability at the 2\% level for
100\,mJy sources and at the 20\% level for 10\,mJy sources on a daily
basis.  This is similar to levels achieved in the (targeted)
MASIV survey (Lovell et al. 2003) which found that more than 50\% of
the AGN in their sample were variable.  With 17 sources per square
degree above 10\,mJy in the sky, \tel\ would usefully 
survey some 350 000 AGN each day.

A small subset of these AGN show variability on timescales of less than a
day, caused by interstellar scintillation; the intra-day 
variable (IDV) sources (Kedziora-Chudczer et al. 1997).
IDVs are of astrophysical
interest because the small angular sizes ($<50\,\mu$as) that they
must possess in order to exhibit interstellar scintillation require them to
have brightness temperatures near or, in many cases, several orders of
magnitude in excess of the $10^{12}\,$K inverse Compton limit for
incoherent synchrotron radiation.
The IDV phenomenon is observed to be highly intermittent in most sources.
What fraction of IDVs are intermittent, and what is the duty
cycle of their IDV?  Is IDV intermittency source- or ISM-related?
The brightening and fading of IDV in intermittent sources, which
\tel\ could automatically measure in hundreds of thousands of objects,
would then address the physics that causes the bright emission in
the first place. IDVs may also provide the
first {\bf direct} detection of the ionized baryons in the IGM at
$1\lesssim z\lesssim6$ through angular broadening induced 
by turbulence in the IGM.
Hints of this are already present in the MASIV data (Lovell et al. 2007),
where reduced variability is seen for sources with $z\gtrsim2$.

Extreme Scattering Events (ESEs) are a type of transient in which the
flux variations are not intrinsic to the source but are caused by
variations in refraction along the line-of-sight (Fiedler et al 1987b;
Romani et al. 1987). In other words ESEs are a lensing
phenomenon; not gravitational lensing, but refraction of radio waves in
ionised gas. It has long been recognised that the lenses which cause ESEs
must be Galactic, probably within a few kiloparsecs, but in the 20 years
since the phenomenon was discovered no satisfactory
physical model has emerged. In part this is a reflection of the
difficulty of explaining the existing data using established ideas about the
interstellar medium. Recently, however, Walker (2007) has suggested that
the lenses must be associated with neutral, self-gravitating gas clouds
and may be the source of the Galaxy's dark matter.
Very little new data on the phenomenon has been obtained since the early
work of Fiedler et al (1987a, 1994) and the field has stagnated. The
all-sky monitoring capability of \tel\ will address this problem 
in a comprehensive way.

\begin{figure}
\begin{center}
\noindent
\includegraphics [width=0.48\textwidth]{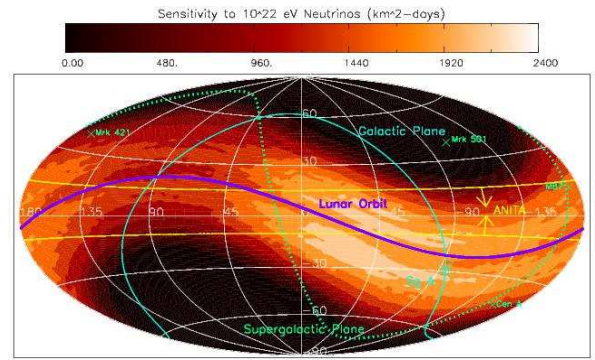}
\end{center}
\caption{Likely sensitivity (km$^2$-days) to $10^{22}$~eV neutrinos \tel\ in
celestial coordinates for one month's observations (40\% duty cycle). A
frequency band of 0.7--1~GHz is assumed. The sky coverage is much broader
than for ANITA Antarctic ice balloon experiment (Miocinovic et al. 2005).}
\label{UHE}
\end{figure}
The origin of the ultra high energy (UHE) cosmic rays (CR), which
have energies extending up to at least $2\times 10^{20}$~eV, is
currently unknown and determining the origin of these particles will
have important astrophysical implications.
A key to untangling the origin
of the UHE CR will be direct detection of UHE neutrinos.
A promising method for the detection of UHE neutrinos is the
Lunar Cherenkov technique, which utilises Earth-based radio
telescopes to detect the coherent Cherenkov radiation (Dagkesamanskii \&
Zheleznykh 1989) emitted when a UHE neutrino interacts in the outer
layers of the Moon (Hankins et al. 1996, James et al. 2007).

Lunar Cherenkov emission produces a linearly polarized,
broadband pulse with sub-nanosecond duration (e.g. Alvarez-Muniz 
et al. 2002, 2006).
In past experiments, false triggering (due to interference)
has proved to be the limiting sensitivity, against which both
multi-antenna coincidence (Gorham et al. 2004) and the
characteristic ionospheric dispersion signature accross a large
bandwidth (James et al. 2007a) have provided powerful
discriminants. The low interference environment of \tel\ means that 
we expect the trigger thresholds --- and hence
sensitivity --- to be limited only by random noise statistics. However,
there are many technical challenges in detecting sub-nanosecond pulses.
\tel\ will serve as the first test-bed
of nanosecond pulse technology which can be directly scaled to
the SKA, though the possibility of a first detection should not
be ruled out.  The predicted sky coverage to UHE neutrinos for
ASKAP is shown in Figure~\ref{UHE}. This includes interesting regions 
where existing limits are weak.

\section{Summary}
\tel\ is a key step on the strategic pathway towards the SKA.
The goals of \tel\ are
to carry out world-class, ground breaking observations,
to demonstrate and prototype technologies for the mid-frequency
SKA and to establish a site for radio astronomy in Western Australia where
observations can be carried out free from the harmful effects of 
radio interference.

In this paper we have outlined the main science themes to be tackled
by \tel, themes which derive from the major issues confronting astrophysics
today such as understanding the evolution, formation
and population of galaxies including our own, understanding the
magnetic Universe, the nature of the transient radio sky and the direct
detection of gravitational waves.

\end{document}